\documentclass[12pt]{article}
\usepackage{graphicx}

\usepackage{relsize}
\usepackage{xspace}
\def\babar{\mbox{\slshape B\kern-0.1em{\smaller A}\kern-0.1em
    B\kern-0.1em{\smaller A\kern-0.2em R}}\xspace}

\newcommand\pubnumber{}
\newcommand\pubdate{\today}

\def\hephy{Institute of High Energy Physics (HEPHY), \\Nikolsdorfer Gasse 18, \\1050 Vienna\\}
\def\support{\footnote{on behalf of the Belle II Collaboration.}}

\def\Title#1{\begin{center} {\Large #1 } \end{center}}
\def\Author#1{\begin{center}{ \sc #1} \end{center}}
\def\Address#1{\begin{center}{ \it #1} \end{center}}

\newcommand\pubblock{\rightline{\begin{tabular}{l} \pubnumber\\
         \pubdate  \end{tabular}}}

\begin{document}
\begin{titlepage}
\pubblock

\vfill
\Title{Dark Sector Physics at the Belle II Experiment}
\vfill
\Author{ Gianluca Inguglia \support}
\Address{\hephy}
\vfill
\begin{abstract}
%
%
The Belle II experiment at the SuperKEKB energy-asymmetric $e^+ e^-$ collider is a substantial upgrade of the B factory facility at the Japanese KEK laboratory. The design luminosity of the machine is $8\times 10^{35}$ cm$^{-2}$s$^{-1}$ and the Belle II experiment aims to record 50 ab$^{-1}$ of data, a factor of 50 more than its predecessor. From February to July 2018, the machine has completed a commissioning run, achieved a peak luminosity of $5.5\times 10^{33}$ cm$^{-2}$s$^{-1}$, and Belle II has recorded a data sample of about 0.5 fb$^{-1}$. Main operation of SuperKEKB has started in March 2019. Already this early data set with specifically designed triggers offers the possibility to search for a large variety of dark sector particles in the GeV mass range complementary to the Large Hadron Collider (LHC) and dedicated low energy experiments; these searches will benefit from more data in the process of being accumulated. This talk will review the state of the dark sector searches at Belle II with a focus on the discovery potential of the early data, and show the first results.

\end{abstract}



\vfill


\end{titlepage}

\setcounter{footnote}{0}

\section{Introduction}
\label{sec:intro}

Despite the success of the standard model (SM) of particle physics in the description of natural phenomena at the fundamental level, there is still information missing that would allow a comprehensive understanding of the Universe. One of the most important issues is dark matter.
Dark matter according to latest cosmological measurements represents a very big fraction in terms of the mass of the known Universe (26$\%$) when compared to ordinary matter (5$\%$). However the only information we have about it comes from gravitational measurements; in fact while dark matter interacts gravitationally with ordinary matter, other interactions have yet to be observed. It becomes then of great importance to fully exploit the potential of new high intensity experimental facilities in the search of signals that can originate from dark matter interactions with ordinary matter, and since this interaction has to be very weak, large amounts of data are required. While one dark matter particle candidate, the neutralino, naturally arises within supersymmetric models, the lack of experimental evidence of its existence suggests that dark matter might be more complex than outlined in those models and it might comprise a new sector of particles, referred to as dark sector, that are neutral under the SM forces but charged under dark forces. In this talk we describe some of the main dark sector searches performed at the Belle II experiment and show specifically the expected discovery potential for \textit{a)} the search of invisible decays of the dark photon ($A'$), \textit{b} the search of axion like particles ($alps, a$), and the search of invisible decays of a low mass $Z'$ boson.

\section{Dark Photon}
\begin{figure}[!ht]
\begin{center}
\resizebox{12.cm}{!}{
\includegraphics{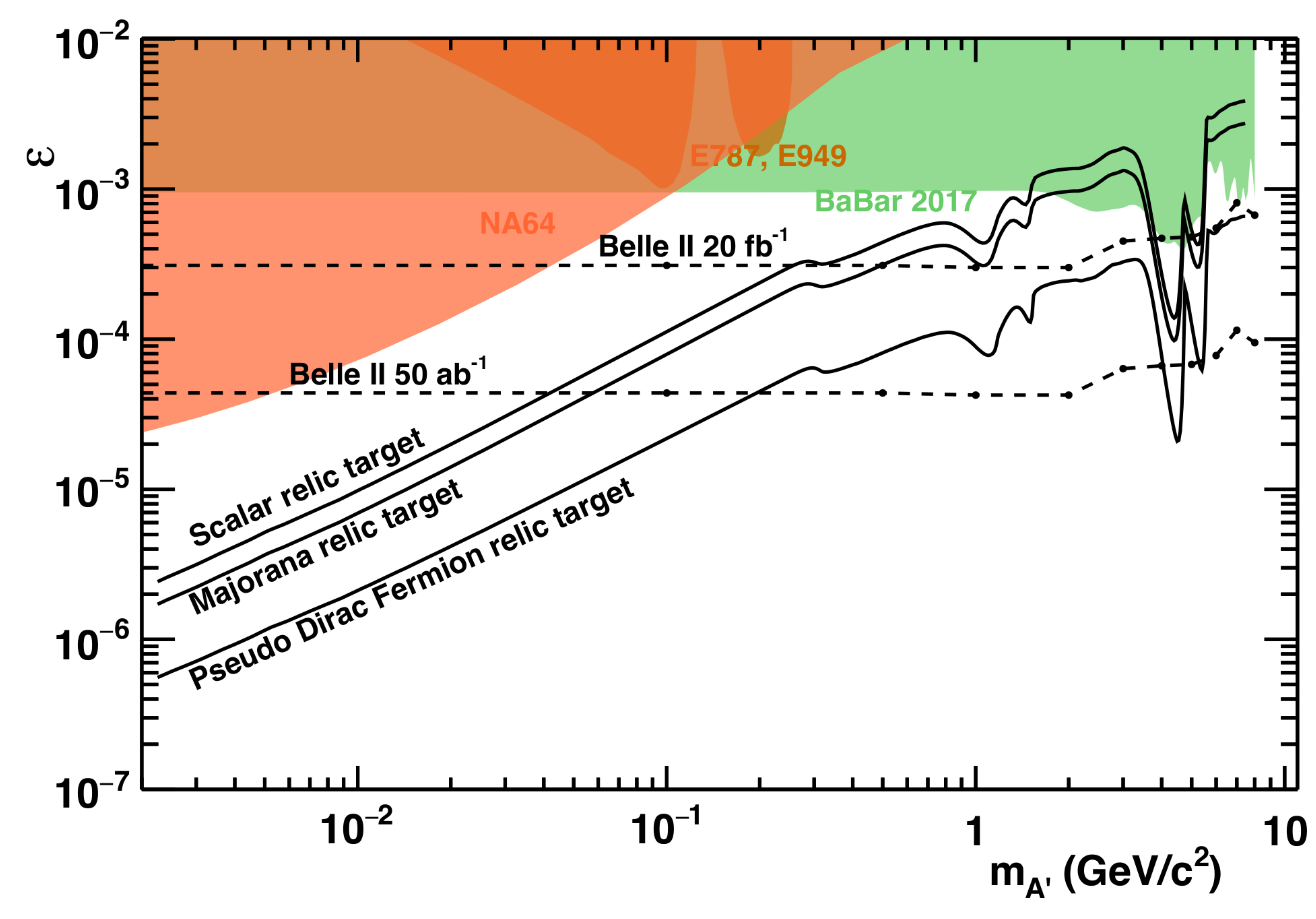}
}
\caption{Expected Belle II sensitivity to the kinetic mixing strength for invisible decays of the dark photon compared to other experiments~\cite{belle2phys}.}\label{fig:figure1}
\end{center}
\end{figure}

The dark photon $A'$ is considered to be the mediator of a hypothetical dark force derived from a $U(1)'$ extension of the symmetry group of the SM. A kinetic mixing between the SM photon $\gamma$ and the dark photon $A'$, $\epsilon F^{Y,\mu\nu}F_{\mu\nu}^D$, with a strength equal to $\varepsilon$ should exist in the interaction Lagrangian and may allow for interactions between SM and dark sector particles. Dark sector or generically dark matter particles would then be neutral under $SU(3)_C \times SU(2)_L \times U(1)_Y$ and charged under $U(1)'_D$ while SM particles would be neutral under $U(1)'$ and the $A'-\gamma$ kinetic mixing term would allow $A'$ to decay to SM particles with very small couplings. Due to the expected low mass of $A'$ (in the range at a few MeV$/$c$^2$ to a few GeV$/$c$^2$~\cite{belle2phys}), $A'$ could be produced in $e^+e^-$ collisions at $B$ and $\tau$-charm factories or in dedicated fixed target experiments in processes that would depend on its mass and its lifetime. At Belle II the dark photon is searched for in the reaction $e^+e^- \to \gamma_{ISR} A'$, with subsequent decays of the dark photon to dark matter $A'\to \chi \bar\chi$ and where $\gamma_{ISR}$ indicates initial state radiation. Such a reaction has a reduced number of particles of the final state that namely consists only of one photon (the ISR photon) plus missing energy. The newly designed single photon trigger will guarantee that these kind of events will be saved for analysis. The signatures for an \textit{on-shell} dark photon production and decays are characterised by the presence of an mono-energetic photon in the final state and, if $s$ is the CM energy of the collision, the energy of the photon is $E_\gamma= \frac{s-M_{A'}^2}{2\sqrt{s}}$, allowing one to precisely estimate the mass of the dark photon. These searches are challenged by the presence of backgrounds that could mimic the invisible dark photon decay signal, and the main contribution to this backgrounds come from QED processes such as $e^+e^- \to e^+e^-\gamma (\gamma)$ (with both the electron and the positron going outside the detector acceptance) and $e^+e^- \to \gamma \gamma (\gamma)$ with one photon going undetected. Preliminary studies have been performed and the sensitivity to the kinetic mixing parameter strength is shown in Fig.~\ref{fig:figure1}~\cite{belle2phys}.

It is interesting to note that also with a small integrated luminosity a very competitive measurement is possible, especially in the region for $M_{A'}>$ few $times$ 10 MeV/c$^2$ where the BABAR experiment starts dominating in terms of sensitivity~\cite{babar}. This is due to the fact that while in the BABAR electromagnetic calorimeter, the crystals were pointing to the interaction region with the consequence that projective cracks between adjacent crystals allowed some particles to escape detection, for example in $e^+e^- \to \gamma \gamma$ with one photon going through a projective crack, and causing an $irreducible$ background. In Belle II, this sort of irreducible background is not present, since the crystals of electromagnetic calorimeter do not point to the interaction region.
\section{Axion-Like Particles}
\begin{figure}[!ht]
\begin{center}
\resizebox{8.cm}{!}{
\includegraphics{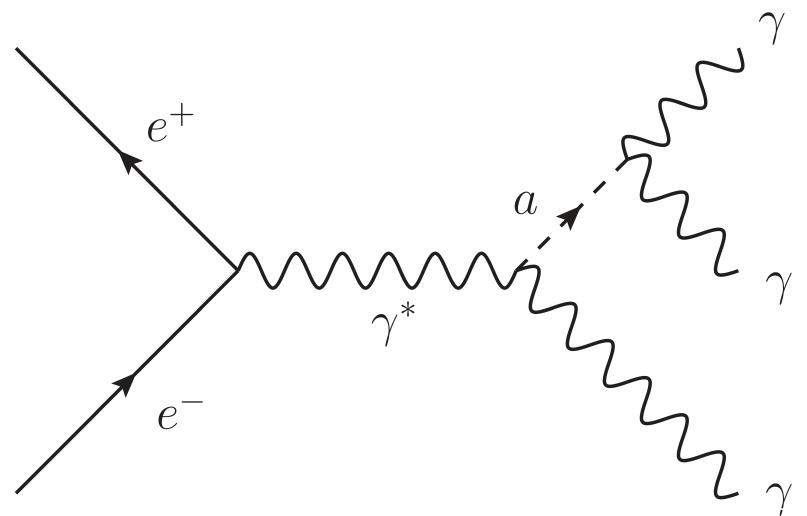}
}
\caption{Feynman diagram showing the ALPS-strahlung process.}\label{fig:figure2}
\end{center}
\end{figure}

\begin{figure}[!ht]
\begin{center}
\resizebox{12.cm}{!}{
\includegraphics{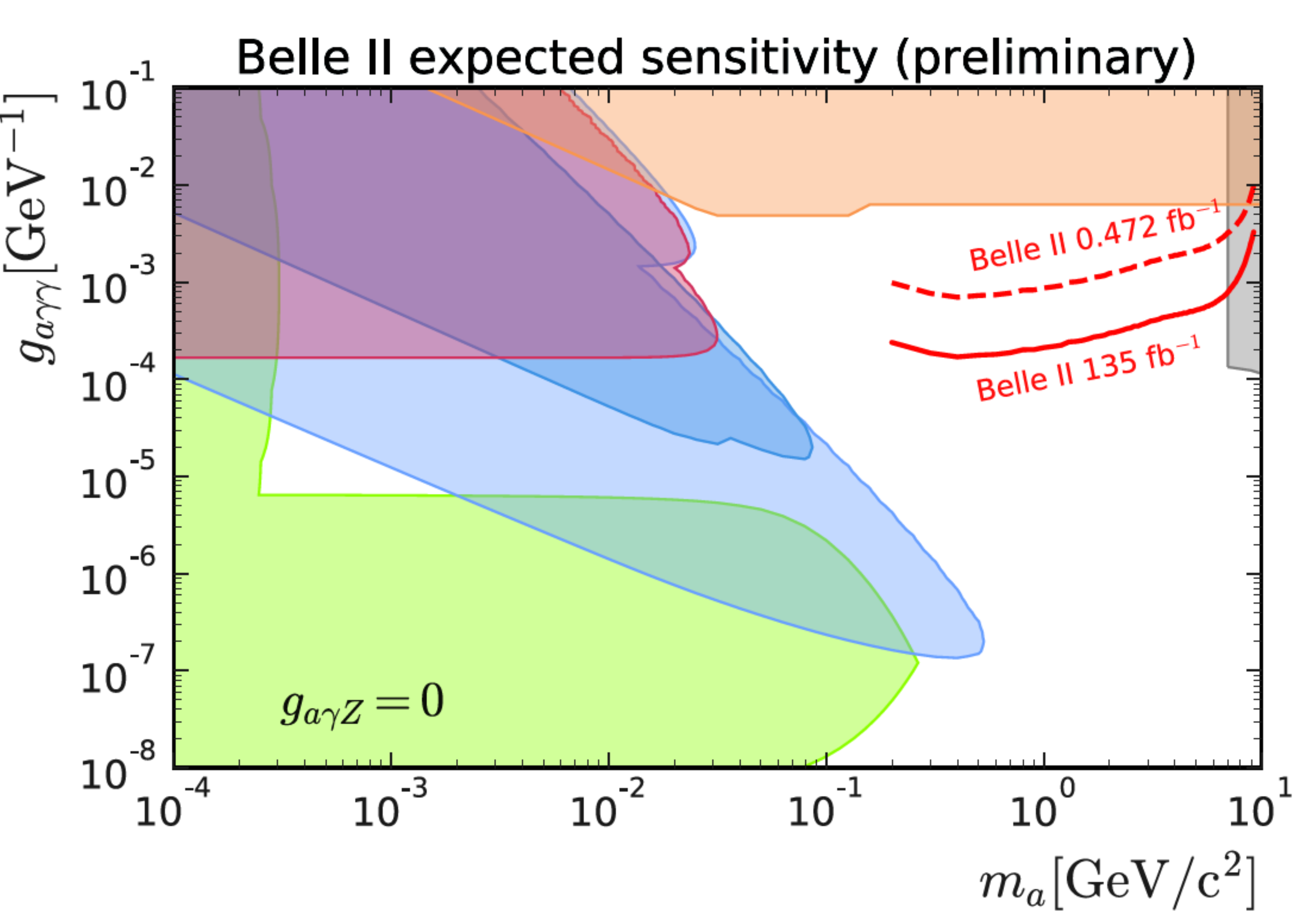}
}
\caption{Expected sensitivity of the Belle II experiment to ALPs parameter space.}\label{fig:figure3}
\end{center}
\end{figure}

With the expression Axion-like particles or ALPs one refers to a class of hypothetical particles similar to the axion in the sense that they are pseudoscalars particles but are not needed to preserve any of the QCD properties that the originally proposed axions have. ALPs can have different couplings to dark sectors, and many different models exist. We consider here the scenario of couplings to SM photons via the Lagrangian term $- \frac{g_{a\gamma\gamma}}{4}aF_{\mu\nu} \tilde{F}^{\mu\nu}$, where $g_{a\gamma\gamma}$ is the coupling constant between the ALPs $a$ and the SM photon, expressed in GeV$^{-1}$.  
One of the simplest searches for ALPS at the Belle II experiment is the so-called ALPS-strahlung process, shown in Fig.~\ref{fig:figure2}.
In the search for ALPs via the ALPS-strahlung process one would expect to observe three photons in the final state, two of which would have an invariant mass equivalent to that of the ALPs $a$ that decayed to the two photons. Properties of the ALPs such as the values of its mass or the coupling to the SM photon will determine its decay length making it visible to the Belle II detector via the decay $a \to \gamma\gamma$ happening inside the detector, or invisible if the decay length is such that the decay will happen outside the detector. We concentrate on the first case, for which a main background component is represented by the QED process $e^+e^- \to \gamma \gamma \gamma$ and it represents the main limitation for this search. The expected sensitivity of the Belle II experiment in the search of ALPs via the ALPS-strahlung process is shown in Fig.~\ref{fig:figure3}~\cite{torben}. Figure~\ref{fig:figure3} also shows that also with just a very little luminosity, such as 135 fb$^{-1}$, it will be possible to uncover for the first time ALPs parameter space with a sensitivity to the $g_{a\gamma\gamma}$ down to 10$^{-4}$.

\section{Low mass dark Z' boson}
\begin{figure}[!ht]
\begin{center}
\resizebox{8.cm}{!}{
\includegraphics{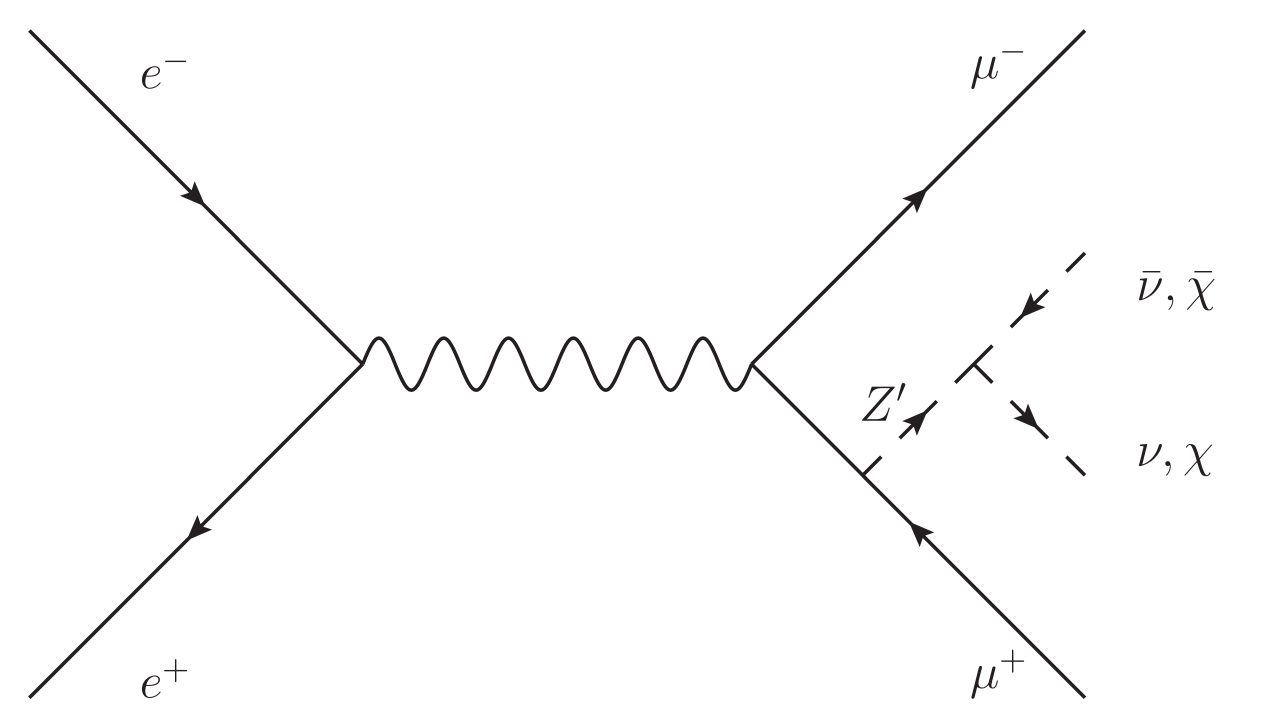}
}
\caption{Feynman diagram showing the production and invisible decay of the $Z'$ boson searched for at the Belle II experiment.}\label{fig:figure4}
\end{center}
\end{figure}

\begin{figure}[!ht]
\begin{center}
\resizebox{12.cm}{!}{
\includegraphics{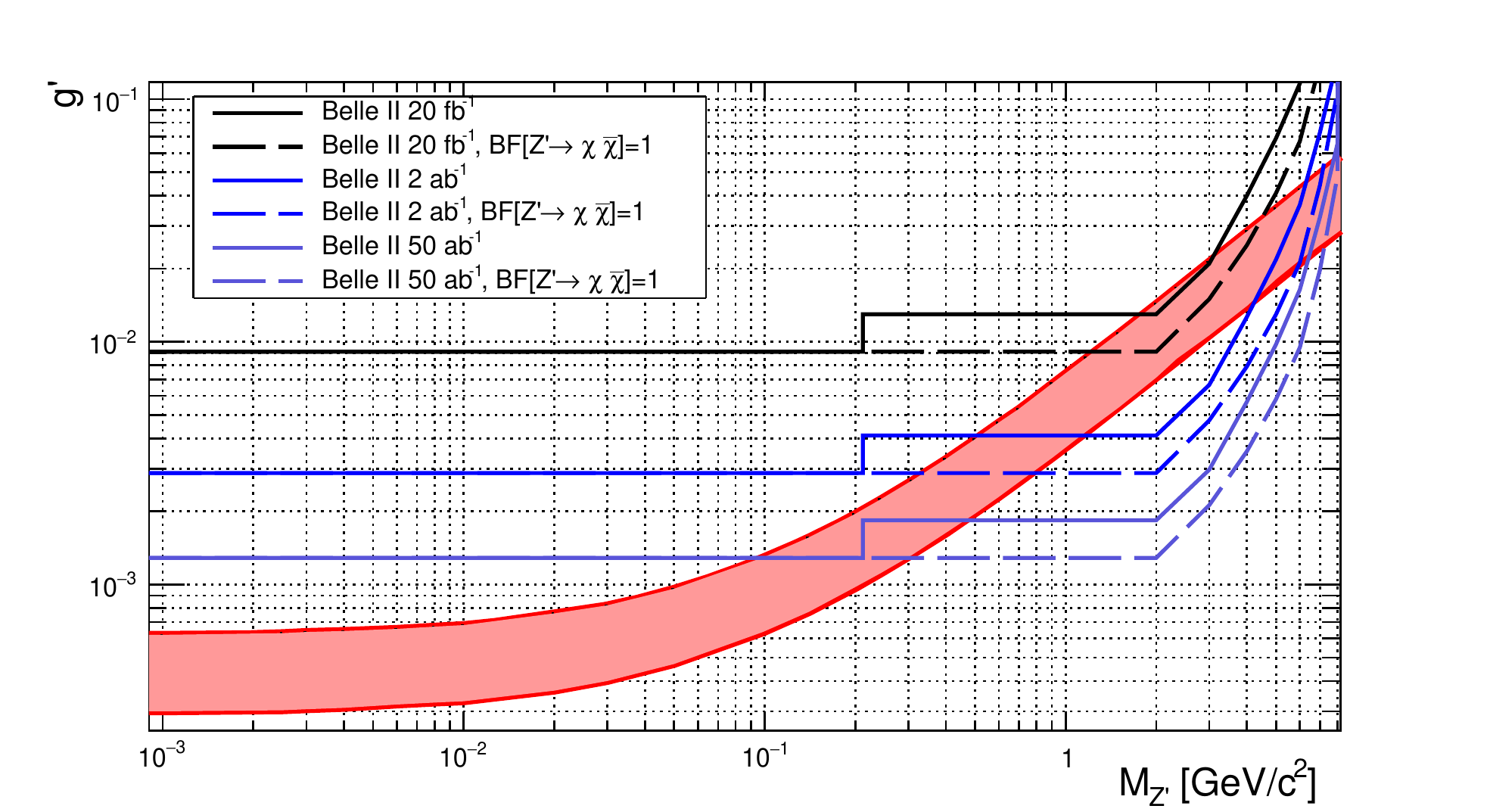}
}
\caption{Expected sensitivity of the Belle II experiment to a $Z'$ parameter space for different luminosity where the solid histograms show the expected sensitivity the Z' decays to neutrino whereas the dashed histogram are calculated assuming an enhancement to the BF[$Z'\to invisible$]=1 as due to the presence of dark matter.}\label{fig:figure5}
\end{center}
\end{figure}

If a dark $Z'$ boson exists and has a mass smaller than $\approx$ 10 GeV/c$^2$, this could be copiously produced in $e^+e^-$ annihilation. We consider here for simplicity the case a low $Z'$ boson belonging to an Abelian symmetry indicated by $L_\mu-L_\tau$. Such a boson would be the mediator of a new force that could create a bridge between SM particles and dark sector particles. This $Z'$ boson would only couple to muons, taus and their respective neutrino species with a new coupling constant indicated with $g'$, so one should search for processes in which these particles are produced so that a $Z'$ boson could be radiated and then decay to any allowed final state: $Z'\to \mu^+ \mu^-$, $Z'\to \tau^+ \tau^-$, and $Z'\to \nu \bar\nu$, where the subscripts indicating the neutrino species have been removed since their flavour can not be identified. A first search for visible decays of the $Z'$ boson to muons was performed by the BABAR experiment~\cite{muonic}. The invisible decay is of interest because the branching fraction can be calculated and if this would be observed and found to be in  disagreement with the predictions, it might be an indication of a decay to new invisible particles, such as dark matter.
Such a search for invisible decays of the $Z'$ boson belonging to the Abelian symmetry $L_\mu-L_\tau$ is being performed for the first time at the Belle II experiment. As process to search for the $Z'$ we consider the QED process $e^+ e^-\to \mu^+ \mu^-$ assuming that one of the muons would then radiate the $Z'$ boson as shown in Fig.~\ref{fig:figure4}, from where it is possible to see that the final state would then consist of two oppositely charged muons plus missing energy.
One very interesting feature of such invisible final state is that it can be constrained and reconstructed by studying the properties of muons and by knowing precisely the energy of the collision of $e^+ e^-$ and properties of the muons in the center-of-mass frame. In fact, one can define the two muon recoil mass as $M_{rec}^{\mu\mu}= s+M_{\mu\mu}^2 - 2 \sqrt{s}E_{\mu\mu}^{CMS}$, which in case of $Z'$ production followed by an invisible decay peaks exactly at the mass of the $Z'$.
After taking into account the efficiency of the trigger, geometrical efficiency and reconstruction efficiency (for a detailed list see~\cite{giacomo}) we estimated a preliminary sensitivity of the Belle II experiment to a $Z'$ as function of its mass and in terms of its coupling constant $g'$. This is shown in Fig.~\ref{fig:figure5} where the sensitivity is also projected to different luminosity and where the solid histograms show the expected sensitivity the Z' decays to neutrino whereas the dashed histogram are calculated assuming an enhancement to the BF[$Z'\to invisible$]=1 as due to the presence of dark matter.
\section{Conclusions}
\label{sec:conclusions}
In this work, the discovery potential of the Belle II experiment when studying dark sector physics has been presented including the expected sensitivities using the early data collected in 2018. The Belle II experiment thanks to the high luminosity and improved detector and analysis techniques is expected to be a major player in the searches for dark sector particles, especially in all those that are characterised by final states with large missing energy in the final state.

\section{Acknowledgments}
This work has been supported by the FWF \textit{Der Wissenshaftsfonds} under the grant number P 311361-N36 ``\textit{Searches for dark matter and dark forces at the Belle II}".

\end{document}